\documentstyle[epsf,twocolumn,aps]{revtex}
\newcommand{\be}{\begin{equation}}
\newcommand{\ee}{\end{equation}}
\newcommand{\bea}{\begin{eqnarray}}
\newcommand{\eea}{\end{eqnarray}}
\newcommand{\lton}{\mathrel{\lower.9ex
                  \hbox{$\stackrel{\displaystyle <}{\sim}$}}}
\newcommand{\pid}{p_{\rm ideal}}
\newcommand{\cid}{c_{\rm QGP}}
\newcommand{\cpi}{c_\pi}
\newcommand{\eid}{e_{\rm ideal}}

\begin{document}                                                
\title{Degrees of Freedom and the Deconfining Phase Transition}
\author{Adrian Dumitru$^{a,b}$ and Robert D.\ Pisarski$^b$}
\bigskip
\address{
a) Department of Physics, Columbia University, New York, New York 10027, USA\\
email: dumitru@quark.phy.bnl.gov\\
b) Department of Physics, Brookhaven National Laboratory,
Upton, New York 11973, USA\\
email: pisarski@quark.phy.bnl.gov\\
}
\date{\today}
\maketitle
\begin{abstract} 
There is a sharp increase in the relative number of degrees of freedom
at the deconfining phase transition.
Characterizing this increase using the Polyakov Loop model, 
we find that for a nearly second order deconfining phase transition,
the medium-induced energy loss turns on rapidly above $T_c$, 
proportional to the relative number of degrees of freedom.
Further, energy loss is logarithmically dependent on 
the screening mass, and thus is sensitive to nearly critical scattering.\\
\end{abstract}

Experiments indicate
that for the central collisions of large nuclei,
$A \sim 200$, there are marked changes between energies of
$\sqrt{s}/A = 17$~GeV, at the SPS, and $130$~GeV, at RHIC \cite{qm01}.  
Comparing central $AA$ collisions to $pp$, 
the spectrum of semi-hard particles is rather different.
At the SPS, in $AA$ the hard $p_t$ spectrum, scaled by the number of binary
collisions, is enhanced over $pp$.
At RHIC, the opposite is true: the semi-hard $p_t$ spectrum
per nucleon-nucleon collision, is suppressed in central $AA$,
relative either to peripheral $AA$, or $p\bar{p}$ \cite{drees}.
This could be the result of ``energy loss'' \cite{wang,eloss2,energyloss},
where a fast colored field loses energy as it passes
through a thermal bath.
In peripheral $AA$ collisions, secondary hadrons are distributed
anisotropically in the transverse momentum $p_t$
~\cite{Ollitrault:1992bk}.  
Experimentally, this azimuthal anisotropy increases with $p_t$ until
$p_t \sim 2$~GeV, at which point it flattens \cite{elliptic}.  This flattening
may also be due to energy loss~\cite{v2_eloss}.

In the limit of infinitely large nuclei, $A \rightarrow \infty$,
it is plausible that the initial energy density produced in 
a central $AA$ collision --- at a fixed value of $\sqrt{s}/A$ ---
evolves into a system in equilibrium at a temperature $T$.
With great optimism, assuming that $A \sim 200$ is near $A = \infty$,
one might imagine that the difference between SPS and RHIC is
because temperatures reached at RHIC exceed $T_c$, the critical temperature
for QCD.  

Thus it is of interest to know how quantities change as one goes through
the phase transition.  In this paper we give an analysis in terms
of the Polyakov Loop model \cite{loop1,loop2,loop3}.

In QCD, there is a large increase in
the number of degrees of freedom at the deconfining phase transition.
We count degrees of freedom as appropriate for the pressure of
free, massless fields at nonzero temperature, 
so if each boson counts as one, then
each fermion counts as $7/8$.
In the hadronic phase, pions contribute 
$\cpi = 3$ ideal degrees of freedom.  
By asymptotic freedom, at infinite temperature QCD with three flavors of
quarks is an Ideal Quark-Gluon Plasma, with
$\cid = 47 \; \frac{1}{2}$ degrees of freedom.  This is an increase of
more than a factor of ten.

To measure the change in the number of degrees of freedom, 
we introduce the relative pressure, $n(T)$: at a temperature $T$, 
this is the ratio of the true pressure,
$p(T)$, to that of an Ideal Quark-Gluon Plasma, 
$\pid = \cid (\pi^2/90)T^4$:
\begin{equation}
n(T) \; \equiv \frac{p(T)}{\pid} \;\;\; . \;\;\;
\label{e1}
\end{equation}
By asymptotic freedom, QCD is an ideal gas at infinite temperature, 
and so
\begin{equation}
n(\infty) = 1 \; .
\end{equation}
For $T < \infty$, corrections
to ideality are determined by the QCD coupling constant,
$\alpha_s \propto 1/\log(T)$, with
$n(T) - 1 \propto - \alpha_s$  \cite{pert}.

For an {\it exact} chiral symmetry which is spontaneously broken by
the vacuum,
about zero temperature the free energy is that of free, massless
pions.  Thus at zero temperature, 
the relative pressure is the ratio of the ideal gas 
coefficients \cite{chiral1}:
\begin{equation}
n(0) \; = \; \frac{\cpi}{\cid} \; . \label{e7b}
\end{equation}
At low temperature, corrections to ideality 
are given by chiral perturbation theory for massless pions, 
$n(T) - n(0) \sim +(T/f_\pi)^4 \; n(0)$, 
with $f_\pi$ the pion decay constant.
In QCD, pions are massive, and the relative pressure is
Boltzmann suppressed at low temperature, 
$n(T)\sim \exp(-m_{\pi}/T)(m_\pi/T)^{5/2}$, so $n(0)=0$.

Given the great disparity between $\cpi$ and $\cid$, consider
an approximation where the hadronic degrees of freedom 
are neglected relative to those of the deconfined
phase \cite{largeN}. 
Then the relative pressure vanishes
throughout the hadronic phase, $n(T) = 0$ for $T < T_c$.  
The question is then: how does the relative pressure
go from zero at $T_c$, when deconfinement occurs, to near one
at higher $T$?

This can be answered by numerical
simulations of Lattice QCD \cite{lattice}.  Consider first
quenched QCD, with pure glue and no dynamical quarks,
which is close to the continuum limit \cite{quenched}.
For three colors, the Lattice finds no measurable pressure 
in the hadronic phase (glueballs are heavy), so our approximation of 
$n(T)=0$ when $T < T_c$ is good.
$n(T)$ increases quickly above $T_c$, 
and is $\sim .8$ by $T \sim 2 T_c$.  To characterize
the change in the relative pressure, consider the
ratio of $e-3p$, where $e(T)$ is the true energy density of QCD,
to the energy of an Ideal Quark-Gluon Plasma, $\eid = 3 \, \pid$:
\begin{equation}
\frac{e - 3p}{\eid} = \frac{T}{3} \frac{\partial n}{\partial T}~.
\label{e2}
\end{equation}
Lattice simulations find that this
ratio has a sharp ``bump'' at $\sim 1.1 T_c$, suggesting
that the relative pressure changes quickly, when the reduced temperature, 
\be
t \equiv \frac{T}{T_c} -1~,
\ee
is small, $t \sim .1$.  

The Lattice is more uncertain with dynamical quarks.  
The pions are too heavy, and it is not near the continuum limit.
So far, the Lattice finds that
$n(T/T_c)$ is about the same with dynamical quarks as without
\cite{lattice,dyn}.
This suggests that the pure glue theory may be a reasonable
guide to how the relative pressure increases above $T_c$.
The approximate
universality of $n(T/T_c)$ is remarkable.  At present,
the Lattice finds no true phase transition in QCD,
with $T_c$ smaller by $\sim .6$ than in the quenched theory \cite{lattice}.
Indeed, even the ideal gas coefficients are very different:
$\cid$ is only $16$ in the quenched theory, versus
$47 \; \frac{1}{2}$ in QCD.  

The greatest change with dynamical quarks is a small, but measurable, pressure
in the hadronic phase.  While in the quenched theory 
$n(T) \sim 0$ for $T< T_c$, with dynamical quarks, although
$n(0) \sim 0$, there is a nonzero relative pressure
at the critical temperature, with $n(T_c)\sim .1$ \cite{lattice}.
Indeed, with no true phase transition, an approximate $T_c$ can only
be defined as the point where the relative pressure increases
sharply, reaching $n\sim .8$ by $2 T_c$ \cite{lattice}.

The Polyakov Loop model \cite{loop1,loop2,loop3} is a mean field
theory for the relative pressure.
In a pure glue theory, the expectation value of
the Polyakov Loop, $\ell_0(T)$, behaves like the
relative pressure: it vanishes when $T< T_c$, and
is nonzero above $T_c$.  Indeed, again by asymptotic freedom,
$\ell_0 \rightarrow 1$ as $T\rightarrow \infty$.
The simplest guess for a potential for the Polyakov Loop is:
\begin{equation}
V(\ell) = - \frac{b_2}{2} |\ell|^2 
+ \frac{1}{4} \left( |\ell|^2 \right)^2 \; .
\label{e3}
\end{equation}
Defining $\ell_0$ as the minimum of $V(\ell)$ for a given $b_2(T)$,
the relative pressure is given by \cite{loop1,loop2,loop3}:
\begin{equation}
n(T) = - 4 V(\ell_0) = \ell_0^4 \; ;
\label{e4}
\end{equation}
$b_2 > 0$ above $T_c$ ($b_2(T)\to 1$ for $t\to\infty$),
and $<0$ below $T_c$.
Thus if the relative pressure changes when the reduced
temperature $t \sim .1$, the change for $\ell_0(T) \sim n^{1/4}$ is
even more rapid, within $2.5\%$ of $T_c$.

For two colors, (\ref{e3}) is a mean field theory for a second order
deconfining transition \cite{second}.
The $\ell$ field is real, and so the potential defines a mass:
$(m_\ell/T)^2 = (1/Z_s) \partial^2 V/\partial\ell^2$, with
\begin{equation}
m_\ell(T)/T \propto \ell_0 \sim n^{1/4} \; ,
\label{e7a}
\end{equation}
where $Z_s$ is the wave function normalization constant for
$\ell$, $Z_s = 3/g^2$, up to corrections of order $g^0$ \cite{wirstam}.
This is measured
from the two point function of Polyakov loops in coordinate space, 
$\propto (1/r)\exp(- m_\ell r)$ as $r\rightarrow \infty$.

For three colors, $\ell$ is a complex valued field,
and a term cubic in $\ell$ appears in $V(\ell)$,
$- b_3(\ell^3 + \ell^*\!^3)/6$.
This produces a first order deconfining transition, where $\ell_0$ jumps
from $0$ at $T_c^-$ to $\ell_c = 2b_3/3$ at $T_c^+$ \cite{loop2}.  
The $\ell$ field has two masses, from its 
real ($m_\ell$) 
and imaginary ($\widetilde{m}_\ell$) parts.
At $T_c^+$, $\sqrt{Z_s} m_\ell/T = \ell_c$; 
from the Lattice, $\sqrt{Z_s} m_\ell/T \sim .3$
\cite{lattice}, which gives $b_3 \sim .45$. 
This small value of $b_3$ reflects the weakly first order
deconfining transition for three colors
\cite{lattice,quenched}.  
The mass for the imaginary part of $\ell$ is
$\sqrt{Z_s} \widetilde{m}_\ell(T)/ T \propto \sqrt{b_3 \ell} \sim n^{1/8}$;
at $T_c^+$, $\widetilde{m}_\ell/m_\ell = 3$.
With dynamical quarks, 
in principle a term linear in $\ell$,
$- b_1(\ell + \ell^*)/2$, can also appear in $V(\ell)$ \cite{banks}. 
If the pion pressure is included below $T_c$, however,
$b_1$ is very small, $\leq .03$. 

Thinking of $\ell_0$ provides a useful way of viewing the
deconfining phase transition.  For a strongly first order
transition --- as appears to occur for four or more colors \cite{four} ---
$\ell_0$, jumps from zero below $T_c$, to a value near one just
above $T_c$.  As $\ell_0$ is near one, the deconfined phase
is presumably well described as a nearly 
Ideal Quark-Gluon Plasma \cite{blaizot}.
In this case, there is a hadronic phase below
$T_c$, and a Quark-Gluon Plasma from $T_c$ immediately on up.

In contrast, for three colors the deconfining transition is weakly
first order.  As the energy density is discontinuous at $T_c$, 
for small $t$ the
relative pressure is linear in the reduced temperature,
\begin{equation}
n(T) \; \sim \; 3 \, r \, t \; ;
\end{equation}
here $r \equiv e(T_c^+)/\eid(T_c)$
is the ratio of the energies at $T_c$, in the deconfined
phase versus an Ideal Quark-Gluon Plasma.
For quenched QCD, $r \sim 1/3$ \cite{quenched}, which gives
$n(T) \sim t$, and so $\ell_0(T) \propto t^{1/4}$.
Except very near $T_c$, this simple
estimate agrees with more complicated analysis using
$b_3\neq 0$ \cite{loop2,loop3}.  For
example, at only $5\%$ above
$T_c$, this estimate gives $\ell_0 \sim .05^{1/4}\sim .5$.
For three colors, then, there is a (non)-Ideal Quark-Gluon Plasma only at
temperatures above $\sim 2 T_c$; between $T_c$ and 
$\sim 2 T_c$, the
Polyakov Loop dominates the free energy, going from 
$\sim .5$ at $1.05 T_c$ to $\sim 1$ by $2 T_c$.

The difference between these two scenarios: a strongly first order
transition, where $\ell_0(T)$ is approximately constant above $T_c$, 
and nearly second order behavior, where $\ell_0(T)$ changes significantly,
is in principle observable.  As an example, consider 
energy loss for a fast parton, with a high energy $E$.  
We first give a general discussion of energy loss in a medium
\cite{eloss2,energyloss}, and then discuss the differences between
a strong first order transition, and one which is nearly second order.

We introduce the energy scale~\cite{energyloss}, 
\begin{equation} \label{Ecritical}
E_{\rm cr} = \frac{m_\ell^2}{\lambda} \; L^2 \; ,
\end{equation}
where $\lambda$ is the mean free path and 
$L$ is the thickness of the medium.  The high-energy jet loses energy by
radiating gluons with energy $\omega < E$. There are several contributions
to the total energy loss of the jet, $\Delta E$, depending on the energy of the
radiation. For the contribution from $\omega>E_{\rm cr}$, which exists if
$E>E_{\rm cr}$, effectively only one
single scattering occurs (this is the so-called factorization regime) and so
that contribution is medium independent~\cite{energyloss}.
In what follows we rather focus on the medium-induced energy loss,
from the region where $\omega$ is less than $E_{\rm cr}$.

For very small frequency, $\omega<E_{\rm LPM}\equiv\lambda m_\ell^2$,
the formation time~\cite{wang,eloss2,energyloss} $t_f\sim \omega/m_\ell^2$
of the radiation from the hard jet is short, and so incoherent radiation
takes place. This is the so-called Bethe-Heitler regime; the contribution to
$\Delta E$ is just a sum from single scatterings on $L/\lambda$ scattering
centers. In the high-energy limit $E,E_{\rm cr}\gg E_{\rm LPM}$ the region of
phase space with $\omega<E_{\rm LPM}$ contributes little to
$\Delta E$ and will be neglected.

The largest contribution is rather from 
the Landau-Pomeranchuk-Migdal (LPM) regime, where
successive scatterings coherently interfere~\cite{wang,eloss2,energyloss}.
Integrating the radiation intensity distribution over $\omega$ from zero
to some energy $E^*$ yields a total energy loss of~\cite{eloss2,energyloss}
\be \label{Eloss*}
-\Delta E \sim \frac{3 \, \alpha_s}{\pi} \; \sqrt{E_{\rm cr} E^*} \;
\log\frac{2E}{Lm_\ell^2}~.
\ee
There is a logarithmic sensitivity to the infrared scale
$m_\ell$~\cite{eloss2,Zakharov:2001iz}.
When the jet energy $E$ is less than the factorization scale $E_{\rm cr}$,
we can integrate $\omega$ all the way up to $E^* =E$, so
\begin{equation}
- \Delta E \; \sim \; 
\frac{3 \, \alpha_s}{\pi} \; \sqrt{E \, E_{\rm cr}} 
\; \log \frac{2E}{Lm_\ell^2}
\;\;\; , \;\;\; (E < E_{\rm cr}) \; .
\label{e5}
\end{equation}
Note that $- \Delta E$ should not exceed $E$.  This requires that
$m_\ell$ is not so small that the logarithm overwhelms the $\sim \alpha_s$.
However, $E_{\rm cr}$ is small near $T_c$, so 
minijets with energies of at least a few GeV are
above $E_{\rm cr}$ anyways, and eq.~(\ref{e5}) does not apply.

Rather, for jet energies greater than $E_{\rm cr}$, 
the total medium-induced
energy loss is given by integrating over $\omega$ up to the factorization
scale $E_{\rm cr}$; setting $E^*=E_{\rm cr}$ in~(\ref{Eloss*}),
\be
- \Delta E \; \sim \; 
\frac{3 \, \alpha_s}{\pi} \; E_{\rm cr} 
\; \log \frac{2E}{L m_\ell^2}
\;\;\; , \;\;\; (E > E_{\rm cr}) \; .
\label{e7c}
\ee

To compute the critical energy $E_{\rm cr}$, we need 
the inverse mean free path, $\lambda^{-1}$.  This 
is approximately the product of the density, $\rho$, 
times the elastic cross section, $\sigma_{\rm el}$.
The elastic cross section is quadratically
divergent in the infrared.  This divergence is naturally
cut off by $m_\ell$, so the elastic cross section
$\sigma_{\rm el} \propto \alpha_s^2/m_\ell^2$, and
\begin{equation}
\frac{m_\ell^2}{\lambda} \propto \rho \; .
\end{equation}
The scale $E_{\rm cr}$ is then proportional to $\rho$;
this follows automatically from our assumption that
$\lambda^{-1} \sim \rho \sigma_{\rm el}$. Thus in the high-energy
regime above $E_{\rm cr}$, ignoring the logarithmic dependence upon
$m_\ell$, energy loss is proportional to $\rho$; below that scale,
to $\sqrt{\rho}$.  This has been emphasized
by Baier, Dokshitzer, Mueller, and Schiff~\cite{energyloss}.

Before giving estimates of $m_\ell$ and $\lambda$, 
we can understand how energy loss changes, depending upon
the order of the deconfining phase transition.
For a nearly second order transition, $m_\ell/T$ is small near $T_c$,
eq.~(\ref{e7a}), and then increases rapidly.
As the energy loss $\Delta E$ depends logarithmically on
$m_\ell$, for small $m_\ell$ energy loss is enhanced.
This is directly analogous to critical opalescence.
In contrast, for a strongly first order transition, $m_\ell$
is large at $T_c^+$, with $m_\ell/T$ approximately constant
with increasing temperature.  

What are reasonable values for $m_\ell$ and $\lambda$~?  
In the extreme perturbative regime, $T \gg T_c$,
(static) electric fields are heavy, with a mass
$\propto \sqrt{\alpha_s} T$, while the static
magnetic fields are light, $m_{mag} \propto  \alpha_s T$.
The inverse
mean free path, $\lambda^{-1}$, equals the damping rate for a gluon
with momentum $\sim T$, and is $\propto  \alpha_s T$.  

At temperatures $\sim 2 T_c$,
this ordering is reversed, as static electric fields are
significantly lighter than static magnetic fields:
$m_\ell \sim 2.5 T$, while the static mass for magnetic
glueballs is $m_{mag} \sim 6 T$ \cite{magnetic}.   
There are no estimates of the damping rate;
we guess that $\gamma \sim \lambda^{-1} \sim T$.  
This seems reasonable for a quasiparticle with such a mass,
as $\gamma/m_\ell \sim 1/2.5 = .4$ is less than one.
If the width were much larger, then
it would not make sense to speak of quasiparticles.
Conversely, in a strongly coupled system, it is unreasonable
to think that the width could be much smaller than the mass.

In the derivation of energy loss, implicitly it is assumed
that multiple scatterings of the hard jet are independent of each other.
This requires that the 
range of the potential is smaller than the mean free path, 
$m_\ell^{-1} < \lambda$ ~\cite{eloss2,energyloss}, which is
equivalent to having quasiparticles with relatively narrow width.

At $2T_c$, then, 
$m_\ell^2/\lambda \sim (2.5 T)^2 \, T$.  
Below $2 T_c$, by our mean field analysis, then,
$m_\ell^2/\lambda \sim 6.25 T^3 n(T)$.  
Notice that as $m_\ell^2 \sim \sqrt{n} \, T^2$, and
$\gamma = \lambda^{-1} \sim \sqrt{n} \, T$, that
$\gamma/m_\ell \sim n^{1/4}$: 
the $\ell$ quasiparticles become narrower as $T \rightarrow T_c^+$.
Consequently, for temperatures near $T_c$, 
\begin{equation} \label{Ecr_scaling}
E_{\rm cr} \sim 20. {\rm GeV} \times n(T) \left(\frac{T}{T_c}\right)^3\, 
\end{equation}
This number is a {\it very} crude estimate, obtained by taking
$L\sim 5$~fm and $T_c \sim 175$~MeV.  For QCD, $n(T_c) \sim .1$,
so then $E_{\rm cr} \sim 2$~GeV.  

It is interesting to note that while the system is becoming increasingly dilute
near $T_c$ for a nearly second order transition, typical minijets with $E$
on the order of several GeV automatically have energies larger than
$E_{\rm cr}$. This implies that i) some part of their energy loss is
medium independent, from the factorization regime $\omega>E_{\rm cr}$;
and that ii) their medium-induced energy loss is given by
eq.~(\ref{e7c}), not~(\ref{e5}). Physically, this is because the formation time
$t_f\sim \omega/m_\ell^2$
of the radiation from the hard jet grows like $n^{-1/2}$ as $T\to
T_c^+$. In all, $E_{\rm cr}$ is small near $T_c$, and the 
regime where $\Delta E$ scales with $\sqrt\rho$ shrinks.
That regime only emerges as $E_{\rm cr}$
increases: by $2T_c$, with $n(2 T_c) \sim 1$, 
$E_{\rm cr}$ has risen dramatically, to
$\sim 160 $~GeV~! Thus, the energy loss~(\ref{e7c}) turns on very rapidly as
$T$ increases from $T_c$.
In part this is because of the factor of $T^3$ 
in eq.~(\ref{Ecr_scaling}), which is present whatever the order
of the phase transition.  
For gauge theories with three colors, such as QCD, however,
{\em additionally} there is an increase 
$\propto n(T)$~\cite{lattice,quenched,dyn}.

With these numbers, the logarithmic sensitivity to the changes
in the screening mass can be significant.  For $E=25$~GeV,
$L=5$~fm, $T=T_c$, then if 
$m_\ell \sim 2.5 T_c$, $\log(2E/(L m_\ell^2)) \sim 2.3$.  
Including the change in the screening mass near $T_c$, 
with $m_\ell \sim 2.5 T_c n^{1/4}$, the logarithm changes to
$\log(2E/(L m_\ell^2)) \sim 3.5$.  This is an increase by about $50\%$.
At smaller $E$, the sensitivity to $m_\ell$ is even stronger.

Even with the limitations of our approximations, it is clear that
since the density vanishes as $T \rightarrow T_c^+$,
{\it any} contribution from the deconfined phase vanishes like
{\it some} power of $n(T)$ \cite{other}.  
For example, estimates indicate that dilepton production from
the (nearly ideal) deconfined phase \cite{dilep} is about as large as that
from the hadronic phase~\cite{dilep2}.  Near $T_c$, dilepton
production from the deconfined phase should be strongly suppressed,
quadratically in the density:
\begin{equation}
\frac{d N_{e^+ e^-}}{d^4x} \propto n^2 \, T^4 \; .
\label{e7}
\end{equation}
This assumes that the density scales as the relative
pressure, $n(T)$.  Whatever the exact form, it is clear that
dileptons from the deconfined phase will turn on later (in $T$,
or $\sqrt{s}/A$) than energy loss, which close to $T_c$ is linear in the
density, and thus in $n(T)$.

The amplitude for dilepton production involves a virtual
photon off the light cone; this has no power like infrared
divergences at either leading or next to leading order in $\alpha_s$,
and so dilepton production should not be especially sensitive 
to the screening mass, $m_\ell$.  This may not be true for
the production of real photons; this 
involves processes on the light cone, with
severe infrared divergences at next to leading order\cite{real}.
At leading order in $\alpha_s$, real photons from
the deconfined phase will be suppressed like $\sim n^2$, as
in eq.~(\ref{e7}).  It is conceivable that infrared effects from higher
order diagrams change this to $\sim n^2/m_\ell^2 \sim n^{3/2}$.
If true, then there is a sequential series of effects: increasing
$T$ from $T_c$, as the relative pressure increases,
first energy loss turns on, then perhaps real photons, then dileptons.

In closing, we remark that changes in the screening mass could also
be significant for the collision rates of quarks and gluons.  This
is particularly true for quantities like the elastic cross
section, the color conductivity, {\it etc}, which are sensitive to $m_\ell$
\cite{Arnold:2000dr}.
\acknowledgements
We thank R.~Baier, K.~Bugaev, F.~Gelis, M.~Gyulassy, L.~McLerran, A.~Peshier,
D.~T.~Son, I.~Vitev, and J.~Wirstam for helpful discussions.
A.D.\ acknowledges support from DOE Grants DE-FG-02-93ER-40764;
R.D.P.\ from DOE Grant DE-AC-02-98CH-10886.

\end{document}